\documentclass[11pt,preprint,flushrt]{aastex}

\usepackage{amsmath}
\usepackage{amssymb}
\usepackage{latexsym}
\usepackage{cancel}
\usepackage{verbatim}
\usepackage{indentfirst}
\usepackage{multirow}
\usepackage{color}
\usepackage{graphicx}
\usepackage{multicol}
\usepackage[colorlinks=true, allcolors=blue]{hyperref}

\usepackage[english]{babel}
\usepackage[utf8x]{inputenc}
\usepackage[T1]{fontenc}

\usepackage[a4paper,top=3cm,bottom=2cm,left=3cm,right=3cm,marginparwidth=1.75cm]{geometry}

\usepackage[colorinlistoftodos]{todonotes}

\begin{document}
\slugcomment{Accepted: March 10 2020, Symmetry, MDPI.}

\title{Image simulations for strong and weak gravitational lensing} 

% Authors, for the paper (add full first names)
\author{Andr\'es A. Plazas}
\email{aplazas@astro.princeton.edu}
\affil{Department of Astrophysical Sciences, Peyton Hall, Princeton University, Princeton, NJ, USA}

\begin{abstract}
{Gravitational lensing has been identified as a powerful tool to address fundamental problems in astrophysics at different scales, ranging from exoplanet identification to dark energy and dark matter characterization in cosmology. Image simulations have played a fundamental role in the realization of the full potential of gravitational lensing by providing a means to address needs such as systematic error characterization, pipeline testing, calibration analyses, code validation, and model development. We present a general overview of the generation and applications of image simulations in strong and weak gravitational lensing.}
\end{abstract}

\maketitle

%\setcounter{secnumdepth}{4}
%%%%%%%%%%%%%%%%%%%%%%%%%%%%%%%%%%%%%%%%%%

\section{Introduction}
Gravitational lensing is defined as the bending of light due to the curvature of space-time caused by any mass-energy distribution, and it is explained by the theory of General Relativity (GR) (\emph{e.g.}, \citet{einstein17,misner1973,carroll04}). 
Lensing phenomena are diverse and depend on the mass-energy distribution of the lens or deflector (\emph{e.g.}, galaxies, cluster of galaxies, the dark matter large scale structure, etc.), as well as on the relative geometric configuration between the source (\emph{e.g.}, distant galaxies, quasars, or even the cosmic microwave background), the lens, and the observer. It is therefore useful and customary to identify several regimes of lensing depending on the particular configuration of the system at hand \citep{dodelson_2017,schneider92}. The case in which multiple images and arcs are produced is known as \emph{strong lensing} \citep{oguri19,treu10}, whereas the less dramatic---but much more common---case, in which the lensing signal is so subtle that it can only be detected statistically, is known as \emph{weak lensing} \citep{bartelmann17,schneider05,bartelmann01}. A third regime known as \emph{microlensing} uses the magnification of the image of a source star when it is aligned with another lensing star along the line of sight. If the lensing star hosts a planet, its lensing effects can also be measured. This technique is used to discover planets at large distances from the Earth \citep{tsapras18,mao12,gaudi12}. In this paper we will focus on image simulations for strong and weak gravitational lensing. 

The astrophysical applications of gravitational lensing are numerous. In cosmology, it is a central technique to test the current concordance model ($\Lambda$CDM, a cosmological constant as a form of dark energy plus cold dark matter) that is supported by several independent lines of evidence (ranging from scales from the cosmic microwave background to galaxies). This successful model, however, calls for the existence of unknown components---dark matter and dark energy---that, together, constitute about 95\% of the contents of the Universe \citep{weinberg13,Huterer_2017,abbott18,planck18}. Strong lensing, in particular, is used to study the distribution of dark matter in galaxies and clusters of galaxies, as well as substructures in dark matter halos, and to probe the distribution of total mass (baryonic and non-baryonic) at small scales \citep{dalal02,vegetti12,nierenberg17,hezaveh16, gilman20,hsueh20}. The number of strong lensing systems and of giant arcs in them can also be used as a cosmological probe, and measurements of time delays of the multiple images of supernovae and quasars can be used to probe the geometry of the universe and its kinematics through cosmography measurements. In general, strong lensing allows for the determination of cosmological parameters such as the Hubble-Lemaître parameter ($H_{0}$) (and to study the current $H_0$ tension, \emph{e.g.}, \citep{Bernal_2016,p2019model,shajib19,wong19}), the dark energy equation of state ($w$), and the total mass density ($\Omega_{\rm{m}}$), and for testing alternatives to the standard cosmological model and GR \citep{jullo10,magana15,caminha16,acebron18,grillo18}. 
Massive clusters also act as cosmic telescopes, and, through strong lensing magnification, allow for the identification and study of distant galaxies from the early universe, black holes, and quasars and active galactic nuclei that otherwise would not be detected with the native instrument resolutions \citep{johnson17a,johnson17b,livermore17}. As probe of both the growth of structure and the expansion history of the Universe, weak lensing of the large-scale structure or \emph{cosmic shear} has been used to investigate the cause of the observed accelerated expansion of the universe (a form of dark energy or models beyond GR) \citep{Aylor_2019, Poulin_2019, Di_Valentino_2017,abbott18,kilbinger15,MUNSHI_2008,hoekstra08,gunn67}).\footnote{The microlensing regime has been used to set constraints on extended dark matter candidates \citep{alcock00} and to identify exoplanets, as mentioned before.} 

While lensing is important for many different applications and its theoretical foundations are well understood, measurement of the signal in each regime is subject to several challenges. Multiple-imaging due to strong lensing is a relatively rare phenomenon as it only affects a small fraction of distant sources \citep{press73}. The measurements usually also require subarcsecond resolutions, and, depending on the applications, knowledge of the redshifts of the source and the deflector. When redshift information is necessary and it is not possible to obtain it via spectroscopy, photometric redshifts are often used as estimates. However, these are difficult to calculate in crowded fields due to the mixing of light from the foreground object with the background source \citep{treu10,jouvel14,molino17}. Moreover, in some cases they might prove to be not accurate enough for the applications at hand \citep{remolina18}. 

Weak gravitational lensing, on the other hand, is fundamentally limited by shape noise---the standard deviation of the intrinsic galaxy shape distribution--- requiring imaging a large number of galaxies to statistically reduce it. In order to address this limitation, several current  (\emph{e.g.}, the Kilo Degree Survey, KiDS\footnote{\url{http://www.astro-wise.org/projects/KIDS/}}), the Hyper Suprime Cam survey (HSC\footnote{\url{http://www.naoj.org/Projects/HSC/HSCProject.html}}), the Dark Energy Survey (DES\footnote{\url{http://www.darkenergysurvey.org/}}))
and future (\emph{e.g.}, the Dark Energy Spectroscopic Instrument \citep[DESI, ][]{DESI16}, the Prime Focus Spectrograph \citep[PFS, ][]{tec14}, the Vera C. Rubin Observatory Legacy Survey of Space and Time (LSST) \citep[LSST\footnote{\url{https://www.lsst.org/}},][]{LSST19}, \emph{Euclid}\footnote{\url{https://sci.esa.int/web/euclid}} \citep{laa11}, and the Wide Field Infrared Survey Telescope, \emph{WFIRST} \footnote{\url{https://wfirst.gsfc.nasa.gov/}} \citep{spergel15}) projects have and are being designed to collect large amounts of data over large fields of view, reaching a point in which systematic errors are comparable or even dominant over statistical errors. Large data sets will also present an opportunity to increase the number of strong lensing systems found (including more complex systems with multiple sources and deflectors). 

Image simulations play a central role in addressing the observational and systematic challenges faced during the measurement of strong and weak gravitational lensing. In general, simulations with known inputs are a fundamental tool for important tasks such as software testing and validation, systematic error characterization, requirement validation, model testing and developing, calibration, analysis testing, etc. Depending on the objective at hand and the computational resources available, there is usually a trade-off between realism and efficiency that results in simulations with varying degrees of complexity (\emph{e.g.}, end-to-end simulations vs simulations where single parameters are changed one at a time). 

The structure of this paper is as follows. Section $\S$\ref{basics} briefly reviews the basic theory of gravitational lensing (for more detailed reviews, see \emph{e.g.}, \citep{schneider05,hoekstra08, Bartelmann_2010,kilbinger15}). Section $\S$\ref{sims} provides an overview of image simulations in for strong and weak gravitational lensing studies.  Simulations of multiple images and giant arcs in strong lensing by galaxy clusters via ray-tracing is explained, as well as how pipelines for strong lensing can be used to simulate observation through any particular instrument. We discuss the use of simulation in assessing lensing inversion codes for mass modeling, and we also point out the increasing use of simulated strong lensing systems as training sets in finding codes that are based on machine learning techniques. We also discuss the role of simulations in redshift distribution estimations, weak lensing systematic errors characterization, and the creation of mock images that include gravitational lensing effects in the context of observations from large astronomical surveys. We conclude in Section $\S$\ref{conclusion}.
%%%%%%%%%%%%%%%%%%%%%%%%%%%%%%%%%%%%%%%%%%
\section{Gravitational lensing basics}
\label{basics}
Photons from background sources travel along null-geodesics in spacetime. When passing a nearby mass density concentration, their trajectories are bent by an amount determined by the \emph{deflection angle}, $\hat{\vec{\alpha}}$. The deflection angle can be derived by considering the trajectory of a photon in a weakly (assuming that $\Phi/c^2$ << 1 and $\dot{\Phi}$ = 0, where $\Phi$ can be thought of as the scalar Newtonian potential that obeys Poisson’s equation) perturbed Friedmann-Lemaître-Robertson-Walker metric and solving the perturbed geodesic equation. Alternatively, it can be derived by using Fermat's Principle and assuming  that light is traveling through a medium with effective refraction index $n =1-2\Phi/c^2$. The deflection angle is thus given by: 
\begin{equation}
    \hat{\vec{\alpha}} = \frac{2}{c^2}\int \vec{\nabla}_{\perp} \Phi ds
\end{equation}
For a point mass---with a gravitational potential given by $\Phi(r)=-GM/r$---the deflection angle is given by $4GM/c^2b$, where $b$ is known as the \emph{impact parameter} or the distance of closest approach to the \emph{lens} or \emph{deflector}. For most situations of astrophysical interest, the spatial scales of the lens itself are much smaller than the angular diameter distances between the source and the deflector ($D_{ds}$) and between the deflector and the observer ($D_{d}$). Under these circumstances, it is possible to use the \emph{thin lens approximation}, where the 3D mass distribution of the deflector is projected to a 2D plane---the \emph{lens plane}---perpendicular to the line-of-sight and characterized by its surface-mass density (with $\vec{\xi}$ as the vector in the lens plane):
\begin{equation}
    \Sigma(\vec{\xi}) = \int \rho(\vec{\xi},s) ds
\end{equation}
The deflection angle at any point $\vec{\xi}$ is given by the sum of the the contributions due to each individual mass element in the plane: 
\begin{equation}
\hat{\vec{\alpha}}=\frac{4 G}{c^{2}} \int \frac{\left(\vec{\xi}-\vec{\xi}^{\prime}\right) \Sigma\left(\vec{\xi}^{\prime}\right)}{\left|\vec{\xi}-\vec{\xi}^{\prime}\right|^{2}} d^{2} \xi^{\prime 2}
\end{equation}
The line-of-sight defines the optical axis between the observer, the lens plane, and the {\em source plane}. For a source at an angular position $\vec{\beta}$ that emits a ray of light with an impact parameter $\vec{\xi}=\vec{\theta} D_{d}$ on the lens plane, the general mapping between $\vec{\beta}$ and $\vec{\theta}$ is given by the the {\em lens equation}: 
\begin{equation}
\vec{\beta}=\vec{\theta}-\frac{D_{d s}}{D_{s}} \hat{\vec{\alpha}}\left(D_{d} \vec{\theta}\right)=\vec{\theta}-\vec{\alpha}(\vec{\theta})
\label{eq:lens}
\end{equation}
The lens equation is, in general, non-linear and can have multiple solutions (this is formally the {\em strong lensing regime}). The second part of Eq. \ref{eq:lens} defines the {\em reduced deflection angle} $\vec{\alpha}$, which can be written as $\vec{\alpha}=\nabla_{\perp}\left(\frac{2 D_{d s}}{D_{s} c^{2}} \int \Phi d s\right)$. The lens equation thus implies that $(\vec{\theta}-\vec{\beta})$ can be expressed as the gradient of a potential $\psi(\vec{\theta})$, known as the {\em lensing potential}, that is a scaled projection of the 3D Newtonian potential $\Phi$ 
\begin{equation}
\psi(\vec{\theta})=\frac{2 D_{d s}}{D_{d} D_{s}} \int \Phi\left(D_{d} \vec{\theta}, s\right) d s
\end{equation}
Defining $\vec{\nabla}_{\theta}=D_{d} \vec{\nabla}_{\perp}$ as the angular gradient, the lens equation can be written as 
\begin{equation}
\vec{\nabla}_{\theta}\left(\frac{1}{2}(\vec{\theta}-\vec{\beta})^{2}-\psi\right)=0
\label{eq:lens2}
\end{equation}
Eq. \ref{eq:lens2} is Fermat's Principle, $\vec{\nabla}_{\theta}t(\vec{\theta}, \vec{\beta})=0$, with the {\em time delay surface} $t(\vec{\theta})$ defined as:
\begin{equation}
t(\vec{\theta}, \vec{\beta})=\frac{1+z_{d}}{c} \frac{D_{d} D_{s}}{D_{d s}}\left(\frac{1}{2}(\vec{\theta}-\vec{\beta})^{2}-\psi\right)
\end{equation}
The time delay surface consists of a geometric term and a gravitational time delay term known as {\em Shapiro delay}. Multiple images are manifested as the stationary points of the surface, and their arrival time difference will depend on the Hubble-Lemaître constant through the angular diameter distances.   For a point mass, the lens equation reads $\beta=\theta-\frac{D_{d s}}{D_{d} D_{s}} \frac{4 G M}{c^{2} \theta}$. When the source and the lens are aligned, the image formed is a ring with an {\em Einstein radius} of 
\begin{equation}
\theta_{E}=\sqrt{\frac{4 G M}{c^{2}} \frac{D_{d s}}{D_{d} D_{s}}}
\end{equation}

The Laplacian of the lensing potential is proportional to the surface-mass density of the lens: 
\begin{equation}
    \vec{\nabla}_{\theta}\psi = \frac{2\Sigma(D_{d}\vec{\theta})}{\Sigma_{c}} \equiv 2 \kappa
\end{equation}
The term $\kappa$ is known as the {\em convergence}, and the {\em critical surface mass density} is defined as $\Sigma_{c} \equiv \frac{c^{2} D_{s}}{4 \pi G D_{d s} D_{d}}$. In the case when the images of the lensed sources are small compared to the spatial scales in which the deflection angle varies considerably, the lens equation can be linearized to obtain local information of the mapping. Its Jacobian is given by by $A_{i j}=\frac{\partial \beta_{i}}{\partial \theta_{j}}=\delta_{i j}-\frac{\partial \alpha_{i}}{\partial \theta_{j}}$, and can be written as 
\begin{equation}
A=\left(\begin{array}{cc}{1-\kappa} & {0} \\ {0} & {1-\kappa}\end{array}\right)+\left(\begin{array}{cc}{-\gamma_{1}} & {\gamma_{2}} \\ {\gamma_{2}} & {\gamma_{1}}\end{array}\right)=(1-\kappa)\left(\begin{array}{cc}{1-g_{1}} & {-g_{2}} \\ {-g_{2}} & {1+g_{1}}\end{array}\right)
\label{eq:jac}
\end{equation}
where the complex shear $\gamma=\gamma_{1}+i \gamma_{2}$ is defined in terms of the derivatives of the lensing potential as $\gamma_{1} \equiv \frac{1}{2}\left(\partial_{1} \partial_{1} \psi-\partial_{2} \partial_{2} \psi\right)$ and $\gamma_{2} \equiv \partial_{1} \partial_{2} \psi$, and the reduced shear $g$ is defined as $\gamma/(1-\kappa)$. The inverse of the Jacobian in Eq. \ref{eq:jac} is known as the {\em magnification tensor}, and its determinant represents the local magnification $\mu$ in the limit of a point source: 
\begin{equation}
\frac{1}{\mu}=
\left|\frac{\mathrm{d}^{2} \beta}{\mathrm{d}^{2} \theta}\right|=|\det \mathrm{A}|
=\left(1-\kappa^{2}\right)-\gamma^{2}
\end{equation}
The curves in the image plane for which the magnification is formally equal to infinity ($\det \mathrm{A} = 0$) are known as {\em critical curves}. The corresponding curves in the source plane are known as {\em caustics}. In practice, the magnification is never infinite and the finite size of extended sources and other optical effects become important \citep{schneider92}. However, if a source lies close to a caustic curve, it will be highly magnified and distorted, producing images such as giant arcs in cluster, for example. 
 
\section{Image simulations}
\label{sims}

\subsection{Lensing by galaxy clusters}

The lens equation in Eq. \ref{eq:lens} is, in general, non-linear, and it will have multiple solutions if the mass of the deflector is large enough and there is a geometrical alignment between the deflector and the background sources. Multiple images and giant arcs can be produced by systems such as galaxy clusters---the largest gravitationally bound objects in the Universe. In order to generate simulated images that include the effects of lensing by systems like these, a mass distribution of the deflector must be produced. For this purpose, analytical models such as a single isothermal sphere or an elliptical power law \citep{Tessore_2015} can be used, with the advantage of being computationally fast, but with the limitation of being too idealized. This limitation can be partially overcome by the use of semi-analytical algorithms such as {\tt{MOKA}} \citep{giocoli12}, in which each component of the system (\emph{e.g}, host dark matter halo, central galaxy, and satellites) is modelled analytically but in which information from state-of-the-art numerical simulations is also used. 

The use of full N-body and hydro-dynamical simulations is an approach to model cluster lenses, with the lensing effects later incorporated through ray-tracing algorithms. In N-body simulations, a box is filled with N massive particles that interact only through gravity (\emph{e.g.}, {\tt{Millennium}} \citep{springel05}, {\tt{Millennium-XXL}} \citep{angulo12}, {\tt{Dark Sky}} \citep{skillman2014dark}, {\tt{OuterRim}} \citep{Habib_2016}, {\tt{EuclidFlagship}} \citep{potter2016pkdgrav3}). Hydro-dynamical simulations, in turn,  attempt to provide a more accurate description of the properties observed in galaxy clusters  by also including physical properties such as gas cooling, heating, and feedback, in addition to dark matter (\emph{e.g.}, {\tt{Illustris}} \citep{Vogelsberger_2014}, {\tt{EAGLE}} \citep{Schaye_2014}, {\tt{MUFASA}} \citep{Dav__2016}, {\tt{APOSTLE}} \citep{Sawala_2016}, {\tt{RomulusC}} \citep{Tremmel_2018}). For a comprehensive review, see \citet{vogelsberger2019cosmological}.

Numerical simulations provide models for the distribution of dark matter and gas in clusters \citep{Gardini_2004,Rasia_2004}. The lensing properties of dark matter halos generated through numerical simulations can be studied by the use of the {\em ray tracing} technique, in which the trajectory of each individual photon is propagated and followed through the system, assuming that the photons emitted from the source are independent \citep{Killedar_2011, meneghetti08,meneghetti10a,plazas19,li16, Metcalf_2014,Petkova_2014}. A large number of light rays is sent through the mass distribution of the lens,  where the deflection of their trajectories is computed and the distorted and magnified images of the background sources are reconstructed. Since each ray is independent, this approach lends itself to computational paralellization. 

Lens models using numerical methods provide information on positions and masses, which can be used to calculate the deflection angle of any light ray that intercepts the lens plane at a given normalized, dimensionless position $\vec{x}\equiv \vec{\xi} / \xi_0$ (for some characteristic distance $\xi_0$ in the lens plane) by summing the contribution from all the possible lens particles in a system with $N$ particles: 
\begin{equation}
\vec{\alpha}(\vec{x})=\sum_{i=1}^{N} m_{i} \frac{\vec{x}-\vec{x}_{i}}{\left|\vec{x}-\vec{x}_{i}\right|^{2}}
\end{equation}
The computational time of this direct approach is of the order of $N^2$. A more efficient approach 
projects the lens particle positions to a regular grid of size $M\times M$, and the mass in each cell is calculated by summing the masses of the particles that belong to that particular cell. A bundle of rays is traced through another regular grid that covers the lens plane, and the deflection angle for each ray ($i$,$j$) is computed by adding up the contributions from each cell ($k$, $l$) in the grid: 
\begin{equation}
\vec{\alpha}_{i j}=\vec{\alpha}\left(\vec{x}_{i j}\right)=\sum_{k=1}^{M} \sum_{l=1}^{M} m_{k l} \frac{\vec{x}_{i j}-\vec{x}_{k l}}{\left|\vec{x}_{i j}-\vec{x}_{k l}\right|^{2}}
\label{sum_alpha}
\end{equation}
Comparing Eq.\ref{sum_alpha} with the definition of convergence, it can be seen that the deflection angle can be written as a convolution of the convergence with the kernel: 
\begin{equation}
\vec{K}(\vec{x}) \equiv \frac{1}{\pi} \frac{\vec{x}}{|\vec{x}|^{2}}
\end{equation}
The deflection angle can be obtained by applying the convolution theorem in Fourier space ($\tilde{\vec{\alpha}}(\vec{k})=2 \pi \tilde{\kappa}(\vec{k}) \tilde{\vec{K}}(\vec{k})$, where the tilde indicates the Fourier transform). Working in Fourier space offers speed advantages (provided that specific boundary and periodicity conditions are met), although the deflection angle can also be obtained by means of direct calculations using tree methods \citep{meneghetti10a,rasia12,Metcalf_2014}. Once the deflection angles are calculated, lensing properties such as shear and magnification can be obtained through the Jacobian of the lens equation (Eq. \ref{eq:jac}).
In the case that multiple deflectors are considered, the lens equation can be generalized to include multiple lens planes, $N_{\mathrm{p}}$: 
\begin{equation}
\vec{\beta}=\vec{\theta}-\sum_{i=0}^{N_{\mathrm{p}}} \frac{D_{i\mathrm{s}}}{D_{\mathrm{s}}} \hat{\vec{\alpha}}^{i}\left(\vec{\theta}^{i}\right)
\label{eq:multiple_lens}
\end{equation}
Analogously, multiple source planes can also be included to provide a more accurate representation of the system at hand. For example, in the case of an isolated lens such as a galaxy cluster the source planes are generated to account for the redshift dependence of the geometric factor $D_{ds}/D_{s}$: source objects are divided into a certain number of redshift bins with centers equally spaced in lensing distance, each one defining a source plane. 

\subsubsection{Observation pipelines}
Using the ray-tracing formalism to include lensing effects, a model of a galaxy cluster or multiple clusters, a given a distribution of sources, and information on observational conditions (\emph{e.g.}, bandpass, detector parameters, integration time, etc.), it is possible to construct an observation pipeline (\emph{e.g}, {\tt{Skylens}} \citep{meneghetti08, meneghetti10a,plazas19}, {\tt{PICS}} \citep{li16}) that produces simulated images of systems observed through any particular instrument. The pipeline {\tt{Skylens}}, for example, generates source galaxies as denoised postage stamps after drawing from the galaxy sample of the \emph{Hubble eXtreme Deep Field} ({\tt{HXDF}} \citep{illingworth13}), and implements the capability to use multiple source and lens planes. Lens models are produced through the use of the semi-analytic code {{\tt MOKA}}, although any analytical dark matter halo model or numerical simulation can be used as well. Input parameters such as exposure time ($t_{\mathrm{exp}}$), the total throughput function ($T(\lambda)$, which includes the quantum efficiency of the detector, the mirror reflectivity, the transmission curve of the filters and the lenses in the optical system, and the total extinction function), sky coordinates, effective telescope diameter ($D$), detector gain ($g$), readout and Poisson noise, and pixel scale ($p$) must be specified by the user in order to prepare a virtual observation through a chosen instrument and telescope, and calculate the measured flux of a galaxy with a given spectral energy distribution. Following the discussion and formulas in \citet{grazian04} (see also \citet{plazas19}), the total photon counts on the detector (in analog-to-digital units, ADU, or digital numbers, DN) from a source with surface brightness
$I(\vec{x},\lambda)$ (erg s$^{-1}$cm$^{-2}$Hz$^{-1}$arcsec$^{-2}$), is given by 
\begin{equation}
    {\rm ADU}_{\rm total}(\vec x) =  \frac{n_\gamma(\vec x)+n_{\rm sky}+n_{\rm dark}}{g}
\end{equation}
where 
\begin{eqnarray}
  n_\gamma(\vec x)& = &\frac{\pi D^2 t_{\rm exp}p^2}{4 h}\int I(\vec
  x,\lambda)\frac{T(\lambda)}{\lambda}{\rm d}\lambda \; \\
  n_{\rm sky} &=& \frac{\pi D^2 t_{\rm exp}p^2}{4 h}\int\frac{T(\lambda)S(\lambda)}{\lambda}{\rm d}\lambda \;
\end{eqnarray}
are the contributions from the source and the sky, respectively (with $h$ as Planck's constant and $S(\lambda)$ is the sky flux per square arcsec), and $n_{\rm dark}$ is the dark current. The zero point of the image, in the AB system, can be calculated as \citep{grazian04}
\begin{equation}
    \mathrm{ZP} = 2.5 \log \left( \frac{\pi D^2 t_{\rm{exp}}}{4 h g}\int \frac{T(\lambda)}{\lambda}{\mathrm {d}}\lambda \right) - 48.6
\end{equation}

\begin{figure}[h]
\centering
\includegraphics[width=0.75\hsize]{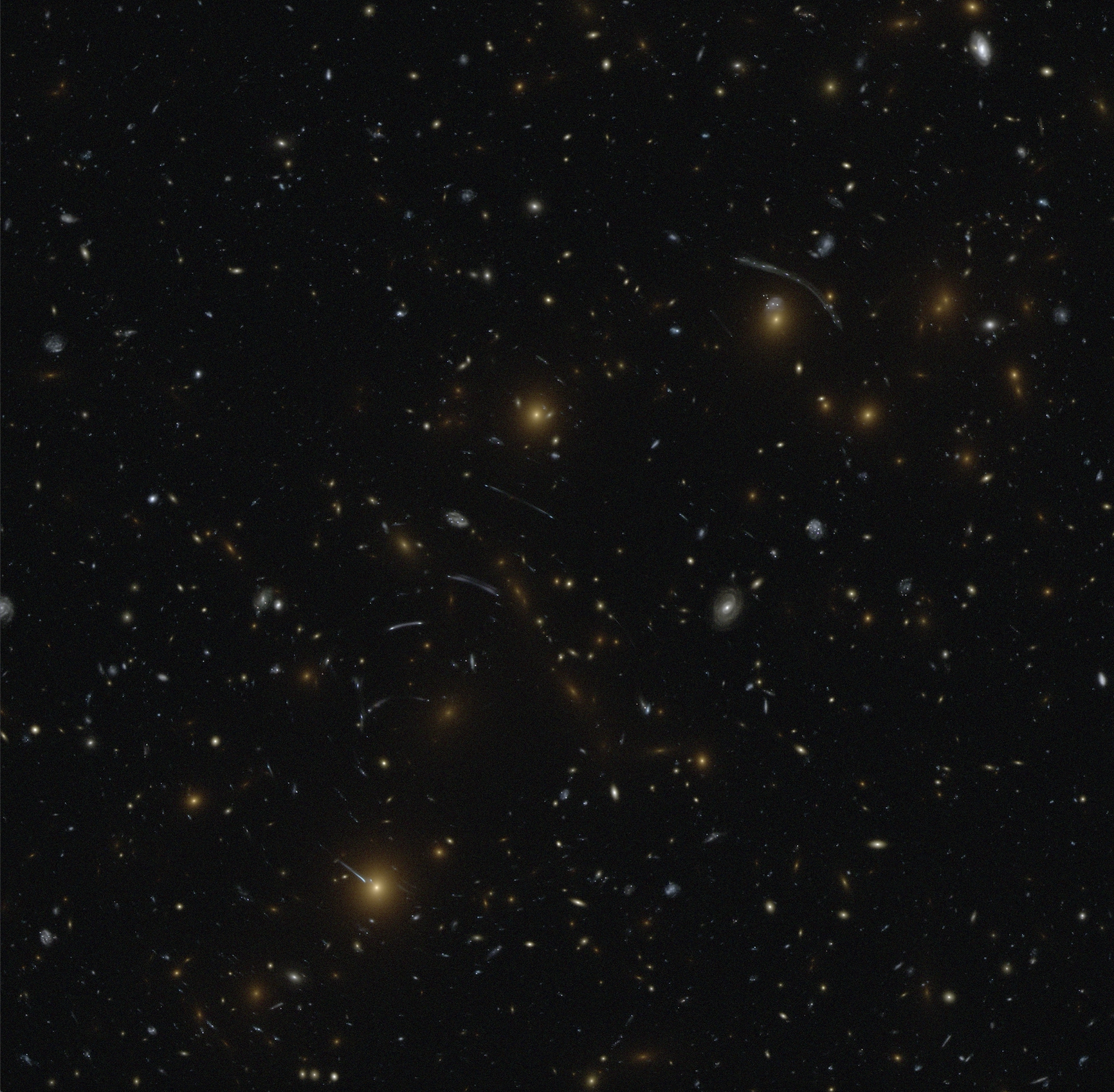}
\caption{Color composite from images produced by the ray-tracing simulation code {\tt{Skylens}}. The image was produced by combining simulated observations taken with the Hubble Space Telescope Advance Camera for Surveys Wide Field Channel in the {\tt{F475W}}, {\tt{F625W}}, and {\tt{F775W}} filters. Taken from \citet{plazas19}}
\label{fig:skylens_hst}
\end{figure}
\subsubsection{Ray tracing}
\label{sec:ray}

Once the telescope, the detector, and the deflector (or deflectors) have been defined, {{\tt Skylens}} accounts for the tidal effect of matter along the line of sight by using Equation (\ref{eq:multiple_lens}) and reconstructs the images of the sources. The images are then convolved by the instrumental PSF, and different sources of noise and sky background are added according to he parameters of the simulated observation. Fig. \ref{fig:skylens_hst} shows an example of an image simulation of the Hubble Space Telescope (HST) Advance Camera for Surveys Wide Field Channel generated by {{\tt Skylens}} through the combination of three different filters. 

\subsection{Image simulations in weak and strong lensing mass modeling}
One of the most important applications of strong and weak lensing by galaxy clusters is constraining of the total mass distribution of the lens, dominated by dark matter \citep{Jullo_2007,Caminha_2019}. Strong lensing observables such as flux ratios, relative image positions, and time delays are used in combination with weak lensing, galactic kinematics, and x-ray information to constrain the matter distributions of clusters via a \emph{lens inversion} process. Knowledge of the matter profile of galaxy clusters is crucial in the understanding of the interplay between dark matter and baryons, as well as the process of large-scale structure formation, and to test the predictions of the standard $\Lambda$CDM cosmological model. High-resolution images of galaxy clusters taken with the Hubble Space Telescope (\emph{e.g.}, {\emph{CLASH}} \citep{postman12}, the HST \emph{Frontier Fields} \citep{Lotz_2017,koek16}) have enabled the community to develop and test different inversion algorithms that, however, do not always result in consistent reconstructions, even when applied to the same systems (see, for example, the analyses of the system {\tt{MACS J1149.5+2223}} by \citet{smith09} and \citet{Zitrin_2009}). The process of mass modeling is subject to many challenges, and in this context, image simulations provide a valuable tool to assess the performance of inversion codes, to find ways to improve them, and to identify the properties of lenses that are mostly impacted by errors during the construction of a lens model (\emph{e.g.}, cluster ellipticities and dynamics, substructures, baryonic physics, etc.). For example, \citet{meneghetti17} use {\tt{MOKA}} and {\tt{Skylens}} to simulate strong lensing observations through galaxy clusters with the characteristics of the HST \emph{Frontier Fields}. The simulations were used to test several lensing inversion methods (parametric, non-parametric or free-form, and hybrid) by different groups.  

Simulations with {\tt{Skylens}} have also been used to test the robustness of estimating cluster masses by using information from weak lensing, and to study hydrostatic biases by comparing with simulated x-rays observations \citep{meneghetti10a,rasia12}.

\subsection{Strong lensing simulations and machine learning methods} 
Image simulations of strong lensing systems have been used to predict that current and future wide-field galaxy surveys (\emph{e.g,} DES, HSC survey, KiDS, LSST, \emph{Euclid}, and \emph{WFIRST}) will produce several to hundreds of thousands of galaxy-galaxy strong lensing systems \citep{collet15}. Many efforts have recently focused on employing techniques from computer vision and machine learning to go beyond traditional approaches such as visual searches of "blue" arcs near "red" galaxies \citep{diehl17}, goodness of fit examinations after fitting a model to all candidates \citep{Marshall_2009, Chan_2015},
and public science challenges \citep{Marshall_2015, More_2015} to discover new strong lensing systems in the large datasets. Neural networks have demonstrated to be able to distinguish between simulated lenses and non-lenses \citep{Lanusse_2017, Hezaveh_2017}. \citet{Jacobs_2019, Jacobs_2019b, Jacobs_2017} have used convolutional neural networks (CNNs, \citep{LeCunBoserDenkerEtAl89}) to produce a catalog of galaxy-galaxy strong lenses (including high-redshift systems) using data from the Dark Energy Survey, and \citet{Petrillo_2017,Petrillo_2019} have correspondingly found hundreds of candidates in KiDS data. \citet{Jacobs_2019b} use the {{\tt LENSPOP}} \citep{collet15} code to generate a training set that consists of hundreds of thousands of labeled simulated examples to train a CNN that classifies lenses and non-lenses. \citet{Metcalf_2019} use N-body ({\tt{Millenium}}) and ray-tracing ({\tt{GLAMER}}, \citet{Metcalf_2014,Petkova_2014}) simulations to analyze a variety of methods including CNN's, visual inspection, and arc finders to assess their efficiency and completeness, and identify biases in the face of large future datasets. 

\begin{figure}
\centering
{\includegraphics[width=0.65\hsize]{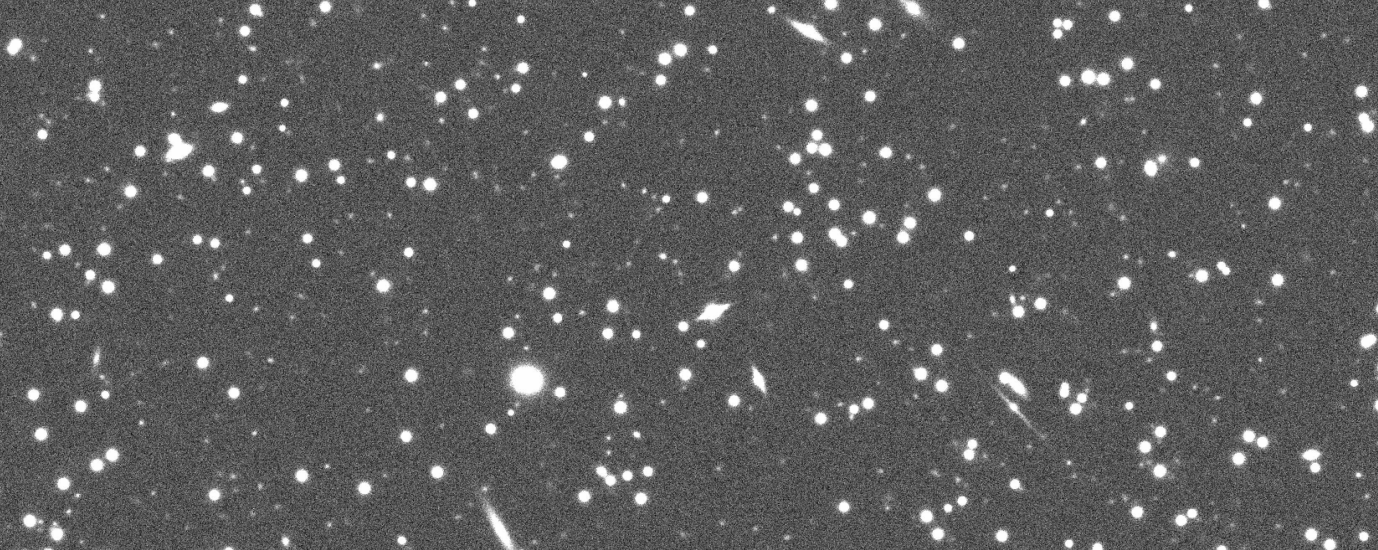}}\hfill
\includegraphics[scale=0.5]{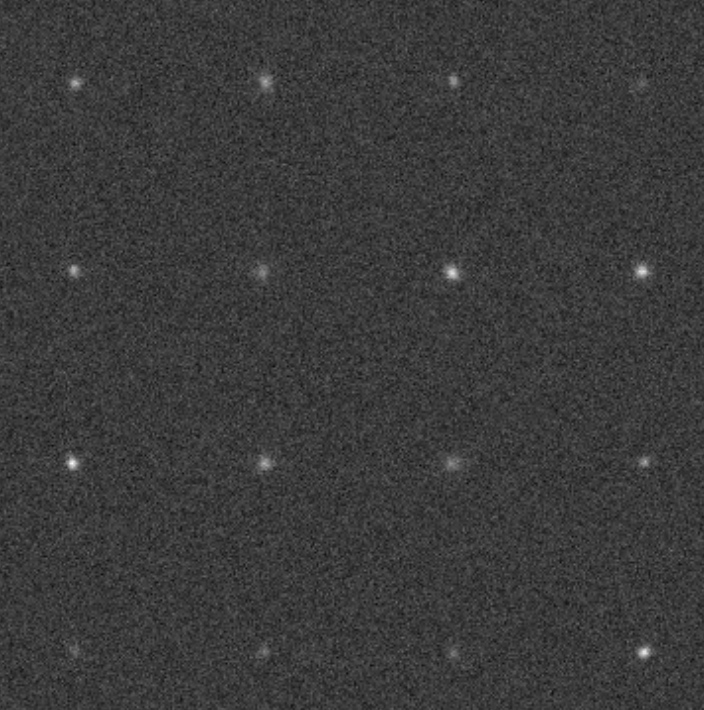}
\caption{Examples of the types of images that can be generated by {{\tt GalSim}}. \emph{Top}: Image of a portion of the focal plane of LSST Camera using realistic parameters, including lensing effects. The image was produced with the publicly available configuration file \url{https://github.com/GalSim-developers/GalSim/blob/master/examples/lsst.yaml}. \emph{Bottom}: Objects with Gaussian surface brightness profiles arranged in a rectangular grid as well-separated postage stamps. }
\label{fig:galsim}
\end{figure}

\subsection{Image simulations for weak lensing systematic errors characterization}

Recent years have seen the rapid advances in astronomical instrumentation and technology enabling the production of large data sets for astronomical investigations. The decrease in statistical uncertainties in these large data sets results in more stringent requirements to control, characterize, and correct for systematic errors. Weak gravitational lensing of the large-scale structure (cosmic shear) plays a central role in these galaxy surveys as a tool to probe the validity of the standard cosmological model and the nature of the dark sector of the Universe. In turn, the understanding and correction of systematic errors lies at the heart of weak gravitational lensing. Systematic errors include shape measurement precision, PSF measurement and deconvolution, instrument signatures, detector (\emph{e.g.}, charge-couple devices and near-infrared sensors) effects, measurement of photometric redshifts, intrinsic alignments, deblending, etc. (for a comprehensive review see \citet{mandelbaum18}). Image simulations with weak gravitational lensing effects have been used to understand these errors and correct for them to the precision required by current and future experiments. End-to-end simulations have been created to bring together the weak lensing community under a common framework and produce challenges such as the \emph{Shear TEsting Programme} ({\tt{STEP}}, \citep{Heymans_2006}) and \emph{GRavitational lEnsing Accuracy Testing} ({\tt{GREAT8}}, \citep{Bridle_2010}). Further iterations of these community-wide challenges ({\tt{GREAT10}} \citep{Kitching_2012} and {\tt{GREAT3}} \citep{Mandelbaum_2014}) focused on simulations in which one parameter a time in the simulation pipeline could be controlled, allowing for a more detailed look at each step in the process of shape galaxy measurement for cosmic shear. The simulations tools to create the images in {\tt{GREAT3}} are publicly available as the modular image simulation code {{\tt GalSim}}\footnote{\url{https://github.com/GalSim-developers/GalSim}} \citep{rowe15}, which has been used extensively to study the impact of systematic effects on weak lensing measurements (\emph{e.g.}, \citep{Plazas_2016,plazas_2017,Plazas_2018,kannawadi15,gruen15, lin2019impact,kamath2019shear}) or as part of pipelines that create mock images where the selection function of the system is directly measured by inserting simulated objects into real data (\emph{e.g.}, {\tt{Balrog}} \citep{Suchyta_2016}, {\tt{SynPipe} \citep{Huang_2017}}). Simulations have also played a central role in the development, testing, and validation of shear estimation methods (\emph{e.g.,} \citep{sheldon19,samuroff18,Bernstein_2016, plazas12}).  In the context of the future \emph{WFIRST} weak lensing program, Troxel \emph{et al.} \citep{troxel2019synthetic} use {\tt{Galsim}} to render images for a simulation suite designed to carefully study weak lensing systematic errors relevant for \emph{WFIRST}'s \emph{High-latitude Imaging Survey} \citep{dor2018wfirst}. Fig. \ref{fig:galsim} shows examples of images produced with {{\tt GalSim}}. 

\subsection{Synthetic sky images and catalogs}
Ray tracing techniques(\ref{sec:ray})---in conjunction with full or approximate N-body simulations (\emph{e.g.}, {\tt COLA} \citep{Tassev_2013, Izard_2017})---are also used in studies of cosmic shear. For example, \citet{derose19} use the {\tt ADDGALS} algorithm to populate dark matter simulations with galaxies, and include gravitational lensing effects via the curved-sky ray-tracing code {{\tt CALCLENS}} \citep{becker13}. They produce a set of 18 synthetic DES Year 1 catalogs to $z = 2.35$ and to a depth of $r \approx 26$ that include galaxy properties (\emph{e.g.}, position, ellipticities, magnitude), photometric errors, and galaxy cluster catalogs (by applying finders such as {\tt{RedMapper}} \citep{Rykoff2014}. These simulations have been used to calculate cosmological observables and tests, including quantities relevant for weak gravitational lensing analyses such as correlation functions (galaxy-galaxy, galaxy-shear, and position-position correlation functions; or 3$\times$2 correlation analyses \citep{abbott18}) and photometric redshift distributions. Simulations such as {\tt{MICE}} \citep{Fosalba2015} use a similar approach to \citet{derose19}, but assume different approximations in the lensing calculation with ray-tracing. Other simulations more focused on weak-lensing statistics using full ray-tracing have also recently been released \citep{Takahashi2017,HarnoisDeraps2013}.

Simulated wide-field images are also used to study systematic errors for weak lensing in a more comprehensive framework that includes data to calibrate the images. \citet{bruderer16, tortorelli2020} use the \emph{Ultra Fast Image Generator} ({\tt{UFIG}}, \citet{berge13})  to produce an implementation of the \emph{Monte Carlo Control Loops} ({\tt{MCCL}}, \citet{refregier14}) framework for weak lensing systematic errors studies and apply it to DES data from science verification (\emph{e.g.}, \citep{Abbott_2016}) data. More recently, \citet{kacprzak19} apply {\tt{MCCL}} to DES year 1 data for cosmic shear studies.  

Alternative approaches to forward-modeling simulation codes such as {\tt{UFIG}} include methodologies that use machine learning methods such as \emph{Generative Adversarial Networks} (GANs) to let the machine infer both the astrophysical and the instrumentation properties of a data set \citep{smith19}

The Dark Energy Science Collaboration of LSST (LSST DESC\footnote{\url{https://lsstdesc.org/}}) also uses wide-field simulations to prepare for the new challenges that the large and complex dataset produced by the LSST of the Rubin Observatory (approximately 40 billion objects and 50 Pb of raw data \citep{juric15,ivezic08}) 
will generate. The latest of these efforts is referred to as \emph{Data Challenge 2} (DC2), the second of three planned synthetic datasets \citep{snchez2020lsst} designed to develop and validate data reduction methodologies and to study the impact of systematic effects on LSST data. DC2 includes the production, validation, and analysis of a 5000 sq-deg mock extragalactic catalogs and 300 sq-deg end-to-end simulation. The extragalactic catalog is produced by the {{\tt CosmoDC2}} pipeline \citep{korytov19}, and includes the N-body simulation {{\tt Outer Rim}}  of 1 trillion particles up to $z=10$, produced with the \emph{Hybrid/Hardware Accelerated Cosmology Code} ({\tt{HACC}}) \citep{Habib_2016}. The simulations include a lensing pipeline that uses particle data from {{\tt Outer Rim}} to generate light cones, project the particles in redshift shells  and uses a ray-tracing algorithm to produce curved-sky lensing maps. The final available information of objects in the catalogs include positions, sizes, shapes, shear, magnification, convergence, magnitudes, etc. The DC2 image simulations uses a subset of the extragalactic catalogs (300 sq-deg) and implements two approaches to produce the images: the Monte Carlo photon shooting code {{\tt PhoSim}} \citep{Peterson_2015} and the code {{\tt ImSim}}\footnote{\url{https://github.com/LSSTDESC/imSim}} which relies on {{\tt GalSim}} to produce the images passing specific LSST information. 

\subsection{Image simulations to assess the accuracy of photometric redshifts}

The determination of accurate galaxy redshift distributions is a key requirement in weak lensing for precision cosmology, especially for the level of requirements needed for the next generation ("stage IV" experiments \citep{albrecht06}) galaxy survey projects that will use weak and strong lensing such as LSST, \emph{Euclid}, and \emph{WFIRST} \citep{Hoyle_2018}.  As such, image simulations constitute a valuable tool to test new algorithms and characterize any biases. \citet{bellagamba12}, for example, use {\tt{Skylens}} to generate mock images of ground- and spaced-base surveys in the different bands in the optical and the near-infrared wavelengths,  and test the performance of template fitting methods of redshift estimation (\emph{e.g.}, BPZ \citep{benitez00}) under different parameters (such as seeing and depth). \citet{Bonnett_2016} use simulations created with {\tt{UFIG}} to calibrate and validate both template fitting and training methods (where a set of galaxies with known spectroscopic methods is used to train machine-learning algorithms, \emph{e.g.}, {\tt{TPZ}} \citep{Carrasco_Kind_2013}, {\tt{ArborZ}} \citep{Gerdes_2010}) in the context of the DES science verification data. 

\section{Conclusion}
\label{conclusion}

We have presented a general overview with some examples of the role that image simulations play in strong and weak gravitational lensing. After 100 years of the formulation of General Relativity, gravitational lensing has positioned itself as one of the most important tools in different areas of astrophysics. In particular, the strong and weak regimes of lensing have become central techniques in current and future experiments that seek to learn more about fundamental problems such as the nature of dark matter and dark energy, the processes of galaxy formation and evolution, the formation of the large-scale structure of the Universe, and the validity of alternative theories of gravity.  Many efforts are focused on the understanding, characterization, and correction of systematic errors and measurement biases in both regimes. Thanks to efforts like these, cosmic shear surveys such as DES have produced the most precise constraints on cosmology from a ground-based survey \citep{abbott18} (with a precision comparable to cosmic microwave background observations), measuring the combination $\sigma_{8}\left(\Omega_{m} / 0.3\right)^{0.5}$ to a precision of about $3.5\%$ \citep{Troxel_2018}. With the advent of stage IV weak lensing surveys with lower statistical errors such as LSST, \emph{WFIRST}, and \emph{Euclid}, the constraints on cosmological parameters will keep improving \citep{Schaan_2017}, but only if systematic errors can be controlled. These wide-field surveys will also increase the number of non-transient (\emph{e.g.}, double source plane lenses) and transient (\emph{e.g.,} lensed quasars and supernovae) strong lensing events, which will help place constraints on the dark energy equation of state parameter and on the current $H_0$ tension. In this context, the creation of simulated images with known inputs is an invaluable tool to achieve the increased required accuracy demanded by these projects and fulfill their scientific potential.

%%%%%%%%%%%%%%%%%%%%%%%%%%%%%%%%%%%%%%%%%%
\vspace{6pt} 

%%%%%%%%%%%%%%%%%%%%%%%%%%%%%%%%%%%%%%%%%%
\section*{Acknowledgments}

We thank Massimo Meneghetti for useful discussions and suggestions to improve earlier versions of this manuscript, and Solný A. Aðalsteinsson for proof-reading its final version. We thank the Department of Physics of Washington University in St. Louis for their hospitality during the preparation of this paper. This research has made use of NASA's Astrophysics Data System Bibliographic Services. 

This paper is dedicated to the memory of Colombian astrophysicist Paul D. Nuñez.  

\bibliographystyle{plainnat}
\bibliography{lensing_sims}

\end{document}